\documentclass{Interspeech}

\interspeechcameraready

\usepackage{balance}
\title{Improving Audio Classification by Transitioning from Zero- to Few-Shot }

\author[affiliation={1}]{James}{Taylor}
\author[affiliation={2}]{Wolfgang}{Mack}

\affiliation{BabbleLabs}{Cisco Systems, Inc.}{Scotland}
\affiliation{BabbleLabs}{Cisco Systems, Inc.}{Germany}
\email{jamesta2@cisco.com, womack@cisco.com}
\keywords{dimensionality reduction, few-shot, audio classification, noise classification, audio embeddings}

\usepackage{comment}
\usepackage{color}
\usepackage{hyperref}
\usepackage{booktabs}
\usepackage{svg}
\usepackage{graphicx}
\usepackage[export]{adjustbox}
\usepackage{multirow}
\usepackage{threeparttable}
\usepackage{pgfplots}
\usetikzlibrary{pgfplots.groupplots}
\usepackage{amsmath}
\usepackage{balance}
\usepackage{tabularx} 

\usepackage[acronym]{glossaries}
\newacronym{clip}{CLIP}{Contrastive Language-Image Pretraining}
\newacronym{llm}{LLMs}{large language models}
\newacronym{lda}{\text{LDA}}{linear discriminant analysis}
\newacronym{mi}{MI}{mutual information}
\newacronym{clap}{CLAP}{Contrastive Language-Audio Pretraining}
\newacronym{map}{mAP}{mean average precision}
\newacronym{mse}{MSE}{mean squared error}
\newacronym{cos}{COS}{cosine similarity}
\newacronym{zs}{ZS}{zero-shot}
\newacronym{fs}{FS}{few-shot}

\begin{document}

\maketitle

\begin{abstract}
State-of-the-art audio classification often employs a zero-shot approach, which involves comparing audio embeddings with embeddings from text describing the respective audio class. These embeddings are usually generated by neural networks trained through contrastive learning to align audio and text representations. Identifying the optimal text description for an audio class is challenging, particularly when the class comprises a wide variety of sounds. This paper examines few-shot methods designed to improve classification accuracy beyond the zero-shot approach. Specifically, audio embeddings are grouped by class and processed to replace the inherently noisy text embeddings. Our results demonstrate that few-shot classification typically outperforms the zero-shot baseline.
\end{abstract}

\section{Introduction}
\label{sec:intro}
Identification and classification of sound events play a vital role in understanding acoustic environments. Competitions like DCASE \cite{Mesaros2019_TASLP} and accompanying datasets like TUT \cite{7760424} and CochlScene \cite{9979822} showcase the importance and interest in acoustic scene classification and event detection. 
Practitioners working on audio machine learning rely on such tools for crucial downstream tasks, such as understanding urban environments and acoustic surveillance in smart cities \cite{10.1145/2647868.2655045, 10761338}, fall alerts in care homes \cite{NHS2020}, monitoring for conservation and ecological research \cite{Browning2017}, and speech enhancement \cite{Sabra2024}. 

Classically, sound event classification is performed using neural networks that map to class activity, which require extensive labeled datasets for each class present in the system \cite{Piczak2015, 7003973}. These models, which need retraining whenever a new class is introduced, are inherently limited in adaptability. In dynamic environments where new events frequently emerge, such constraints become increasingly apparent. To address these challenges, few-shot learning approaches have explored how to classify with minimal labeled examples, often leveraging classical machine learning techniques to reduce data dependence and increase generalization. While \cite{9414406} explores the power of pre-training for a specific domain, \cite{9413584, singh24b_interspeech} demonstrate how classes can be added incrementally and continually. In \cite{8901732, 10446908}, the authors explore how few-shot approaches perform in multi-label datasets. Recent exploration into lightweight models achieve robust performance with limited labeled data, providing a foundation for more adaptable systems.

Alternatively, inspired by the multi-modality of \acrfull{clip} \cite{pmlr-v139-radford21a}, early attempts to constrastively model text and audio \cite{Wu2022, Guzhov2022} modeled audio and its labels in the same space. Further models such as CLAP showed that by leveraging mass-amounts of audio and natural language descriptions, contrastive models can be successfully used to classify, retrieve, and even caption audio \cite{Elizalde2023, niizumi24_interspeech, Wu2023a, Elizalde2024}. Bringing the robustness of language modeling to the problem allowed sound event classification to be approached as a zero-shot problem. These models, while successfully modeling audio and their descriptors together, still show room to improve in terms of understanding sound sequences, sound concurrency, and disambiguating foreground and background sounds \cite{Wu2023}.
In the latest rendition of audio-language models, Pengi framed all problems as text generation problems and has allowed for enhanced captioning, question answering, source-distance discrimination, sentiment/emotion recognition, and music analysis \cite{Deshmukh2023}. Meanwhile, Audio-Flamingo shows augmenting \acrlong{llm} to understand audio has added-on the ability to count sound occurrences, describe the ordering of sounds, and assess sound quality \cite{Kong2024}. 

While these models have become increasingly flexible and impressive, they inherently rely on text embeddings. These text embeddings are general purpose, but also introduce variability, sometimes leaving practitioners to try and find the `best' prompt. In an effort to de-noise these embeddings from other tasks and make them most applicable to such a scenario, we explore light-weight approaches to few-shot, closed-set audio classification. Our approach side-steps the use of these text embeddings by defining classes with a few audio samples instead. In this way, we harness the power of pre-trained audio encoders and remain adaptable to new classes.

Our contribution is a practical and effective extension of modern zero-shot, text-based audio classification tools to a few-shot setting using audio embeddings, showing consistent improvements with minimal data. Another contribution is a rigorous
evaluation of the proposed method and baselines. Our empirical results demonstrate that this approach significantly improves on existing methods, offering a more adaptable solution for audio classification in dynamic environments. 

The paper is structured as follows. In Section~2, we introduce a signal model and embedding-based zero-shot audio classification. The proposed method is presented in Section~3. Section~4 contains information about the hyperparameters and the data we used. Finally, we evaluate the performance in Section~5 followed by a brief conclusion in Section~6.\\

\section{Fundamentals}
\label{sec:fundamentals}

We consider a dataset $\mathcal{X} = \{x_n \mid n = 1, \ldots, N\}$, consisting of \( N \) discrete audio signals \( x_n \in \mathbb{R}^{T_n} \), where \( T_n \in \mathbb{N} \) specifies the number of samples in each signal. Correspondingly, we define a set of labels \( \mathcal{Y} = \{y_n \mid n = 1, \dots, N\} \), with \( y_n \in \mathcal{C} \) representing the class label of the \( n \)-th audio recording, and \( \mathcal{C} \) is the set of possible classes. To handle the variability in audio lengths and to extract meaningful representations, we employ an embedding-extractor model \( \mathcal{F}_A \colon \mathbb{R}^{T_n} \to \mathbb{R}^L \). This model maps each audio recording \( x_n \) to a fixed-dimensional embedding \( e_n \in \mathbb{R}^L \) of length \( L \in \mathbb{N} \), defined as
\begin{equation}
e_n = \mathcal{F}_A(x_n),
\end{equation}
resulting in the set of embeddings $\mathcal{E_A} = \{e_n \mid n = 1, \ldots, N\}$. Similarly, we define a set of text embeddings \( E_T \) using an embedding-extractor \( \mathcal{F}_T \), which processes text akin to \( \mathcal{F}_A \). The training of \( \mathcal{F}_A \) and \( \mathcal{F}_T \) involves reducing the distance between paired audio-text embeddings in a contrastive manner. During inference, each audio class is described textually and transformed via \( \mathcal{F}_T \) to yield a reference embedding \( e_{c} \in \mathbb{R}^L \) for class \( c \). Audio samples are classified by transforming them using \( \mathcal{F}_A \) and mapping them to the closest text embedding in a zero-shot manner \cite{Elizalde2024}. The challenge lies in selecting text that yields the `optimal' embedding for classifying a specific audio class, due to the complexity and variability of language. Different texts may describe the same audio class in various ways, making it difficult to determine which description will best facilitate classification.

\section{Few-Shot Audio Classification Methods}
\label{sec:proposed_method}

\begin{table*}[ht]
\centering
\caption{ Top shows zero-shot baselines, bottom shows FS methods. For ESC-50 and FSD50K, results are obtained from the respective papers. For BBL, results are obtained using open-source implementations. For methods that use feature selection via \acrshort{mi}, we use the best $K$ for a given method/dataset pair (see Figure~\ref{fig:MI_Combined}).}

\begin{tabularx}{\textwidth}{@{\hspace{20pt}}c@{\hspace{20pt}}c@{\hspace{30pt}}c@{\hspace{30pt}}c@{}}
\toprule
Model (ZS)       & \textbf{BBL (dev/eval, acc)}                                 & \textbf{ESC-50 (5-fold, acc)}                                                       & \textbf{FSD50K (dev/eval, mAP)}                              \\ \midrule 
CLAP22      & 0.602                               & 0.826                                                        & 0.302                               \\ 
PENGI       & 0.547                               & 0.920                                                        & 0.468                               \\ 
CLAP23      & 0.623                               & \begin{tabular}[c]{@{}c@{}}0.948 (FT: \textbf{0.983})\end{tabular} & 0.485                               \\ \midrule 
            Model (FS) & \textit{$|\mathcal{E}^c|$=10  \quad   $|\mathcal{E}^c|$=20  \quad   $|\mathcal{E}^c|$=50} & \textit{$|\mathcal{E}^c|$=20}                                                & \textit{$|\mathcal{E}^c|$=10   \quad   $|\mathcal{E}^c|$=20   \quad   $|\mathcal{E}^c|$=50} \\ \midrule
$\text{FS}^{|\mathcal{E}^c|}_{\text{AVG}}$    & 0.678   \qquad   0.700    \qquad   \textbf{ 0.716   }  & 0.970   & 0.540   \qquad     0.561    \qquad    0.579   \\ 
\addlinespace[2pt]
$\text{FS}^{|\mathcal{E}^c|}_{\text{LDA}}$    & 0.615    \qquad    0.623    \qquad    0.632   & 0.968   & 0.440   \qquad     0.487    \qquad    0.526   \\ 
\addlinespace[2pt]
$\text{FS}^{|\mathcal{E}^c|}_{\text{MI+AVG}}$ & 0.681    \qquad    0.698    \qquad   \textbf{0.716}    & \textbf{0.971}   & 0.559   \qquad     0.576    \qquad    \textbf{0.586}   \\
\addlinespace[2pt]
$\text{FS}^{|\mathcal{E}^c|}_{\text{MI+LDA}}$ & 0.563   \qquad     0.612   \qquad     0.650   & 0.969   & 0.509   \qquad     0.551    \qquad    0.559   \\ \bottomrule
\end{tabularx}
\label{tab:results}
\end{table*}

In many practical scenarios, a small number of audio samples for all classes already exists or can be gathered with minimal effort. Extracting audio embeddings from such a development set and constructing $e_c$ from them in a few-shot manner instead of text leads to an improved classification, as we can see in Section~\ref{sec:performance}. Few-shot classification is advantageous over zero-shot classification here, as $e_c$ is directly obtained in a way that optimizes the assignment of embeddings to classes. In the following, we present different few-shot methods to compute $e_c$. We define the set of audio embeddings per class $c$ used to compute $e_c$ as $\mathcal{E}^c$. 
\begin{itemize}
\item \textbf{Averaged Embeddings}: In this method, we compute the average of the embeddings for each class \( c \) using 
\begin{equation}
e_c = \sum_{e_i \in \mathcal{E}^c} w_i \cdot e_i,
\label{equ:mean}
\end{equation}
where \( w_i \) denotes the weight associated with embedding \( e_i \). Two different weighting schemes are employed. The first scheme assigns \( w_i = \frac{1}{|\mathcal{E}^c|} \), where \( |\mathcal{E}^c| \) is the number of embeddings in the set, effectively making  (\ref{equ:mean}) the arithmetic mean. The second scheme assigns weights as \( w_i = \frac{1}{|\mathcal{E}^c| \cdot \lVert e_i \rVert_2 } \), where \( \lVert e_i \rVert_2 \) denotes the \( l_2 \)-norm of \( e_i \), resulting in Equation (\ref{equ:mean}) representing the average of normalized embeddings. This normalization aligns with the properties of cosine similarity, where the direction of vectors, rather than their magnitude, is critical. This approach is consistent with the training methodology of CLAP, where cosine similarity was used.

Subsequently, an audio embedding \( e_n \) is assigned to a class \( c \) based on a minimum distance criterion $d$ given by
\begin{equation}
c = \arg\min_{c \in \mathcal{C}} d(e_c, e_{n}).
\end{equation}
We employ two distance metrics for this classification: \acrfull{mse} and \acrfull{cos}, the latter of which is consistent with the training process of CLAP. 
    
    \item \textbf{\acrlong{lda} and \acrlong{mi}}: We train a \acrfull{lda} model on the audio embeddings $\{\mathcal{E}^c \forall c \in C\}$ classes to obtain a classification system. During inference, the trained \acrshort{lda} model is used to estimate $c$ from $e_n$. In Section~\ref{sec:performance}, we observe a reduced performance of \acrshort{lda} compared to simple averaging. We assume the curse of dimensionality \cite{bellman1957dynamic} to be a reason for the reduced performance. Hence, we additionally employ dimensionality reduction \cite{10.1016/j.cosrev.2021.100378} of $e_n$ before \acrshort{lda} using the \acrfull{mi} approach \cite{298224, vergara2014review}. 

\end{itemize}

\section{Experimental Setup}
\label{sec:data}
\subsection{Data}
The performance evaluation is based on three datasets: BBL (internal), ESC-50, and FSD50K. These datasets provide a diverse range of sound events and evaluation scenarios. \textbf{BBL} consists of 3600 audio signals divided equally among 24 humanly annotated sound classes. The audio originates from AudioSet \cite{7952261}. The classes cover a wide range of indoor and outdoor sounds, such as cars, wind, vacuum cleaners, as well as human non-speech sounds like baby crying or laughter. The 150 signals per class are split into non-overlapping sets of 100 for development and 50 for evaluation. \textbf{ESC-50} \cite{10.1145/2733373.2806390} contains 2000 audio signals, each 5 seconds long, distributed equally among 50 sound event classes. The classes are organized into five broad topics: animals, natural sounds, human sounds, interior/domestic, and exterior/urban noises. Each audio sample is assigned a single label. The dataset uses a 5-fold cross-validation split for evaluation. \textbf{FSD50K} \cite{Fonseca2022} includes more than 50,000 audio signals distributed across 200 sound classes. The audio is organized hierarchically based on the AudioSet ontology, leading to multi-label annotations where a single recording may belong to multiple related categories. The classes cover a wide array of everyday sounds, such as musical instruments, animals, household sounds, and crowd chatter. Recordings vary in length and quality, reflecting real-world variability. The dataset is provided with a development and evaluation split, and performance is typically reported using \acrfull{map}.
\subsection{Few-Shot Methods}
To obtain $\mathcal{E}^c$ for the \acrfull{fs} methods, we choose $|\mathcal{E}^c|$ training samples from the dev set. In the case of FSD50K, we specifically choose those with the least class overlap with other files. We then preprocess audio using the pretrained audio encoder of CLAP (Version 2023) for each dataset separately. To obtain $e_c$, the embeddings are subsequently either averaged per class (AVG), used to train an LDA system, or dimensionality reduced using mutual information (MI) plus subsequently AVG or LDA. We refer to the different \acrshort{fs} methods as $\text{FS}^{|E^c|}_{\bullet}$ where $\bullet \in \{\text{\text{AVG}}, \text{\text{LDA}}, \text{\text{MI+AVG}}, \text{\text{MI+LDA}}\}$ represents the respective method. When using \acrshort{mi}, we vary the number of features selected by \acrshort{mi} a $K$ ratio of $|\mathcal{C}|$ with $K \in \{1/2, 1, 2, 4, 8, 16, 32\}$ such that the embedding size after MI is $|\mathcal{C}|\cdot K$. We decided to vary $K$ in this way in order to create a fair comparison between the datasets, as we believe $|\mathcal{C}|$ and the number of features needed to differentiate between classes are closely linked.

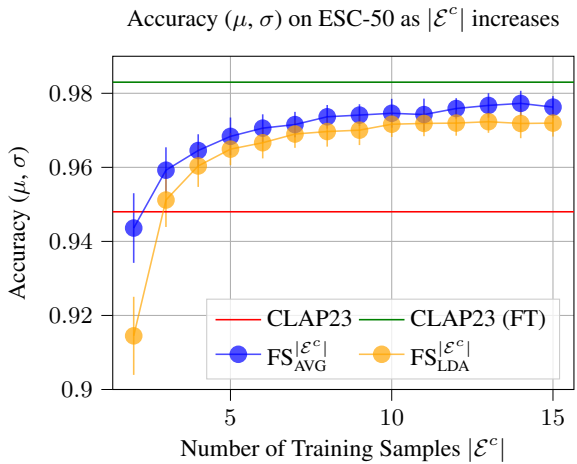
\begin{figure}[t]
    \hspace{-5mm}
\begin{tikzpicture}

\definecolor{darkgray176}{RGB}{176,176,176}
\definecolor{green}{RGB}{0,128,0}
\definecolor{lightgray204}{RGB}{204,204,204}
\definecolor{orange}{RGB}{255,165,0}

\begin{axis}[
legend cell align={left},
height=6cm, 
every mark/.append style={mark size=0.5},
width=0.45\textwidth,
legend columns=2,
legend style={
  fill opacity=0.8,
  draw opacity=1,
  text opacity=1,
  at={(0.97,0.03)},
  anchor=south east,
  draw=lightgray204
},
tick align=outside,
tick pos=left,
title={Accuracy (\(\displaystyle \mu\), \(\displaystyle \sigma\)) on ESC-50 as \(\displaystyle |\mathcal{E}^c|\) increases},
x grid style={darkgray176},
xlabel={Number of Training Samples \(\displaystyle |\mathcal{E}^c|\)},
xmajorgrids,
xmin=1.35, xmax=15.65,
xtick style={color=black},
y grid style={darkgray176},
ylabel={Accuracy (\(\displaystyle \mu\), \(\displaystyle \sigma\))},
ymajorgrids,
ymin=0.9, ymax=0.99,
ytick style={color=black}
]
\path [draw=blue, draw opacity=0.7, semithick]
(axis cs:2,0.934200039430261)
--(axis cs:2,0.953008891989197);

\path [draw=blue, draw opacity=0.7, semithick]
(axis cs:3,0.953049769985872)
--(axis cs:3,0.965410508474407);

\path [draw=blue, draw opacity=0.7, semithick]
(axis cs:4,0.960157227289665)
--(axis cs:4,0.968933681801244);

\path [draw=blue, draw opacity=0.7, semithick]
(axis cs:5,0.963270430767998)
--(axis cs:5,0.973456841959275);

\path [draw=blue, draw opacity=0.7, semithick]
(axis cs:6,0.966856651947771)
--(axis cs:6,0.974319818640465);

\path [draw=blue, draw opacity=0.7, semithick]
(axis cs:7,0.968117195305033)
--(axis cs:7,0.974986569616914);

\path [draw=blue, draw opacity=0.7, semithick]
(axis cs:8,0.970471681607659)
--(axis cs:8,0.976838924452947);

\path [draw=blue, draw opacity=0.7, semithick]
(axis cs:9,0.971187752641668)
--(axis cs:9,0.977042941395478);

\path [draw=blue, draw opacity=0.7, semithick]
(axis cs:10,0.970707123044262)
--(axis cs:10,0.978504998167859);

\path [draw=blue, draw opacity=0.7, semithick]
(axis cs:11,0.969942075955512)
--(axis cs:11,0.978563671170925);

\path [draw=blue, draw opacity=0.7, semithick]
(axis cs:12,0.973095825425038)
--(axis cs:12,0.978679066349853);

\path [draw=blue, draw opacity=0.7, semithick]
(axis cs:13,0.973440190133789)
--(axis cs:13,0.980005376645112);

\path [draw=blue, draw opacity=0.7, semithick]
(axis cs:14,0.973868825298597)
--(axis cs:14,0.980676629246857);

\path [draw=blue, draw opacity=0.7, semithick]
(axis cs:15,0.973288496935188)
--(axis cs:15,0.979244836398145);

\path [draw=orange, draw opacity=0.7, semithick]
(axis cs:2,0.903944114114029)
--(axis cs:2,0.925038342026322);

\path [draw=orange, draw opacity=0.7, semithick]
(axis cs:3,0.943891802694329)
--(axis cs:3,0.958450539648013);

\path [draw=orange, draw opacity=0.7, semithick]
(axis cs:4,0.954715092538774)
--(axis cs:4,0.966025648201966);

\path [draw=orange, draw opacity=0.7, semithick]
(axis cs:5,0.96058096284669)
--(axis cs:5,0.969247608581882);

\path [draw=orange, draw opacity=0.7, semithick]
(axis cs:6,0.962412990373429)
--(axis cs:6,0.97088112727363);

\path [draw=orange, draw opacity=0.7, semithick]
(axis cs:7,0.965270038302409)
--(axis cs:7,0.972709759677389);

\path [draw=orange, draw opacity=0.7, semithick]
(axis cs:8,0.965591419571227)
--(axis cs:8,0.973741913762106);

\path [draw=orange, draw opacity=0.7, semithick]
(axis cs:9,0.966011627335004)
--(axis cs:9,0.974074394170372);

\path [draw=orange, draw opacity=0.7, semithick]
(axis cs:10,0.968207247768714)
--(axis cs:10,0.975081641120175);

\path [draw=orange, draw opacity=0.7, semithick]
(axis cs:11,0.968402776219591)
--(axis cs:11,0.975321361711444);

\path [draw=orange, draw opacity=0.7, semithick]
(axis cs:12,0.968547036346899)
--(axis cs:12,0.975357725557863);

\path [draw=orange, draw opacity=0.7, semithick]
(axis cs:13,0.969324368108918)
--(axis cs:13,0.975317607199724);

\path [draw=orange, draw opacity=0.7, semithick]
(axis cs:14,0.96787088513307)
--(axis cs:14,0.975821422559237);

\path [draw=orange, draw opacity=0.7, semithick]
(axis cs:15,0.968310209294138)
--(axis cs:15,0.975583124039196);

\addplot [semithick, red]
table {%
1.35 0.948
15.65 0.948
};
\addlegendentry{CLAP23}
\addplot [semithick, green]
table {%
1.35 0.983
15.65 0.983
};
\addlegendentry{CLAP23 (FT)}
\addplot [semithick, blue, opacity=0.7, mark=*, mark size=3, mark options={solid}]
table {%
2 0.943604465709729
3 0.959230139230139
4 0.964545454545454
5 0.968363636363636
6 0.970588235294118
7 0.971551882460973
8 0.973655303030303
9 0.974115347018573
10 0.974606060606061
11 0.974252873563218
12 0.975887445887446
13 0.97672278338945
14 0.977272727272727
15 0.976266666666667
};
\addlegendentry{$\text{FS}^{|\mathcal{E}^c|}_{\text{AVG}}$}
\addplot [semithick, orange, opacity=0.7, mark=*, mark size=3, mark options={solid}]
table {%
2 0.914491228070175
3 0.951171171171171
4 0.96037037037037
5 0.964914285714286
6 0.966647058823529
7 0.968989898989899
8 0.969666666666667
9 0.970043010752688
10 0.971644444444445
11 0.971862068965517
12 0.971952380952381
13 0.972320987654321
14 0.971846153846154
15 0.971946666666667
};
\addlegendentry{$\text{FS}^{|\mathcal{E}^c|}_{\text{LDA}}$}
\end{axis}

\end{tikzpicture}
    \caption{Mean and standard deviation of $\text{FS}^{|\mathcal{E}^c|}_\text{\text{AVG}}$ and $\text{FS}^{|\mathcal{E}^c|}_\text{\text{LDA}}$ using between 2 and 15 training samples, using 30 runs for each $|\mathcal{E}^c|$ value. CLAP23 (FT) is the fine-tuned CLAP model for that specific data set.}
    \label{fig:ESC50_var}
\end{figure}

\begin{figure*}[t]
  \includegraphics[width=\linewidth]{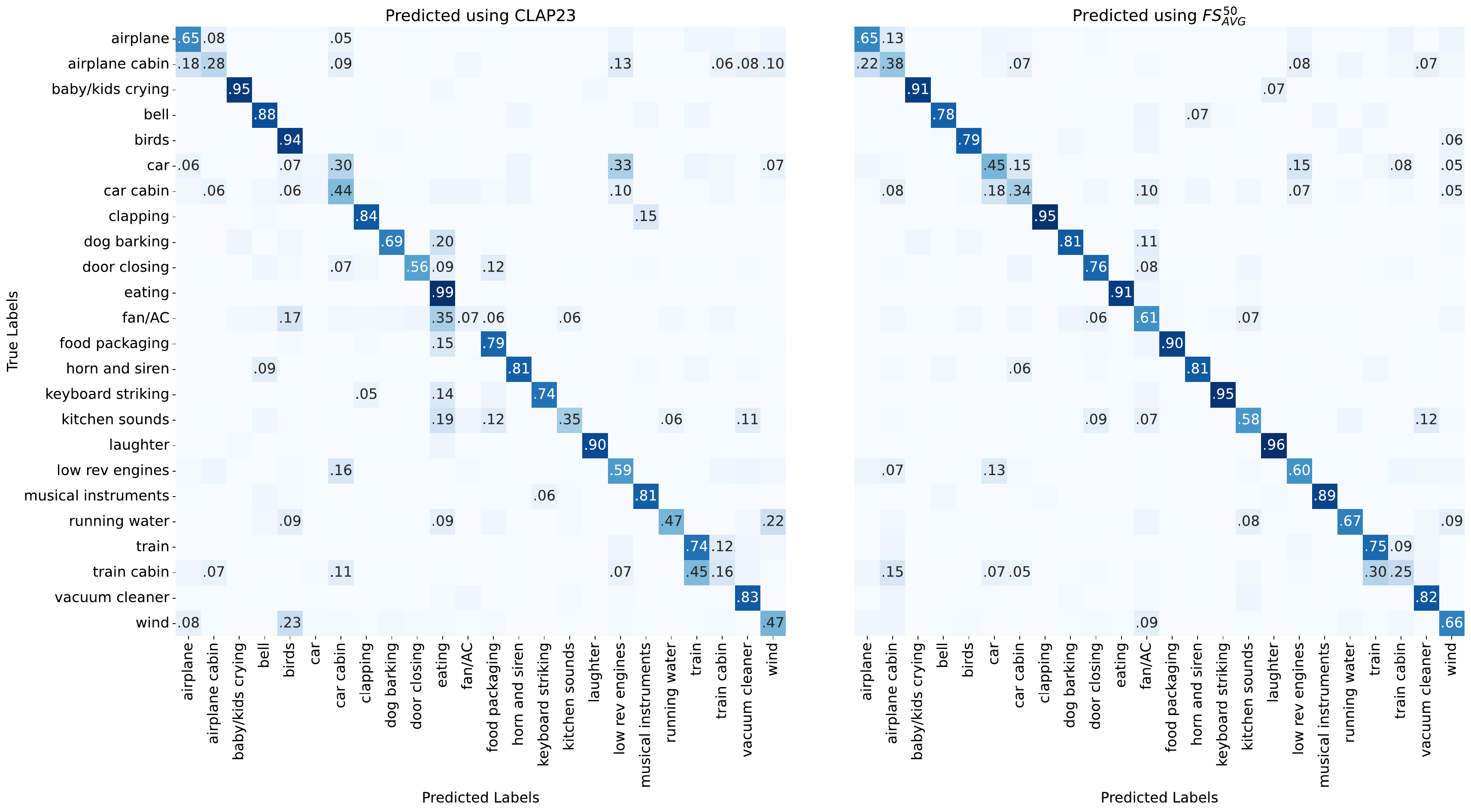}
  \caption{Confusion matrix for BBL. On the left, we see the comparison between the true label and using base CLAP23. On the right, we see a much stronger diagonal trend and an improvement in most classes when using $\text{FS}^{|\mathcal{E}^c|}_{\text{AVG}}$}
    \label{fig:Confusion_Matrices}

\end{figure*}

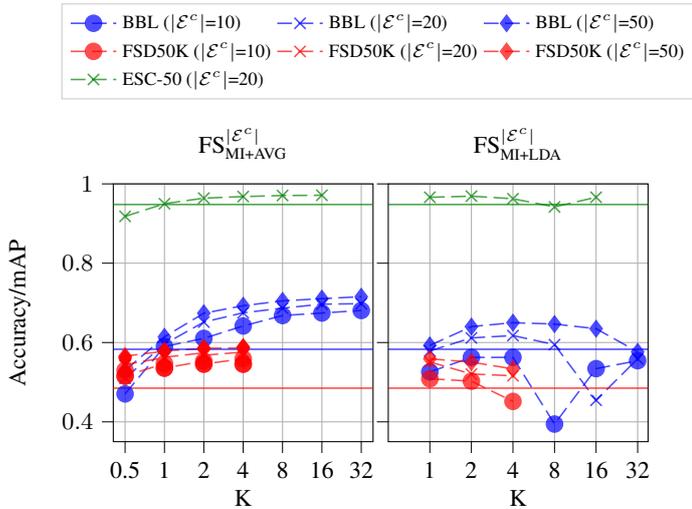
\begin{figure}[h!]
    \hspace{-10mm}
\begin{tikzpicture}

\definecolor{darkgray176}{RGB}{176,176,176}
\definecolor{green}{RGB}{0,128,0}
\definecolor{lightgray204}{RGB}{204,204,204}

\begin{groupplot}[
        group style={
            group size=2 by 1,      
            horizontal sep=.2cm,     
            vertical sep=1cm        
        },
        width=5cm,                 
        height=5cm                 
    ]
\nextgroupplot[
legend cell align={left},
legend columns=3,   
legend style={
  fill opacity=0.8,
  draw opacity=1,
  text opacity=1,
  font=\scriptsize,  
  at={(-0.2,1.7)},
  anchor=north west,
  draw=lightgray204
},
tick align=outside,
tick pos=left,
title={$\text{FS}^{|\mathcal{E}^c|}_{\text{MI+AVG}}$},
x grid style={darkgray176},
xlabel={K},
xmajorgrids,
xmin=-0.3, xmax=6.3,
xtick style={color=black},
xtick={0,1,2,3,4,5,6},
xticklabels={0.5,1,2,4,8,16,32},
y grid style={darkgray176},
ylabel={Accuracy/mAP},
ymajorgrids,
ymin=0.35, ymax=1,
ytick style={color=black}
]
\addplot [semithick, blue, opacity=0.7, dash pattern=on 5.55pt off 2.4pt, mark=*, mark size=3, mark options={solid}]
table {%
0 0.4705
1 0.59
2 0.6105
3 0.64166
4 0.66805
5 0.6744
6 0.6808
};
\addlegendentry{BBL ($|\mathcal{E}^c|$=10)}
\addplot [semithick, blue, opacity=0.7, dash pattern=on 5.55pt off 2.4pt, mark=x, mark size=3, mark options={solid}]
table {%
0 0.5101
1 0.5961
2 0.6515
3 0.675
4 0.6866
5 0.6967
6 0.6979
};
\addlegendentry{BBL ($|\mathcal{E}^c|$=20)}
\addplot [semithick, blue, opacity=0.7, dash pattern=on 5.55pt off 2.4pt, mark=diamond*, mark size=3, mark options={solid}]
table {%
0 0.5295
1 0.61439
2 0.67386
3 0.6928
4 0.705
5 0.7106
6 0.7155
};
\addlegendentry{BBL ($|\mathcal{E}^c|$=50)}
\addplot [semithick, red, opacity=0.7, dash pattern=on 5.55pt off 2.4pt, mark=*, mark size=3, mark options={solid}]
table {%
0 0.53
0 0.515124207198336
0 0.515549972383381
1 0.547
1 0.535181459179818
1 0.535721353475322
2 0.553
2 0.545472153961443
2 0.545830185080078
3 0.559
3 0.545538406715711
3 0.545491823722299
};
\addlegendentry{FSD50K ($|\mathcal{E}^c|$=10)}
\addplot [semithick, red, opacity=0.7, dash pattern=on 5.55pt off 2.4pt, mark=x, mark size=3, mark options={solid}]
table {%
0 0.545
0 0.541118846207789
0 0.544408875741207
1 0.562
1 0.562500379871989
1 0.562887626787781
2 0.574
2 0.567821845025537
2 0.567802365577913
3 0.576
3 0.567819037408308
3 0.568058909984402
};
\addlegendentry{FSD50K ($|\mathcal{E}^c|$=20)}
\addplot [semithick, red, opacity=0.7, dash pattern=on 5.55pt off 2.4pt, mark=diamond*, mark size=3, mark options={solid}]
table {%
0 0.56
0 0.56318906262681
0 0.5658179795311
1 0.578
1 0.578471779811893
1 0.578975313216577
2 0.585
2 0.585071465146605
2 0.58559358779736
3 0.586
3 0.588299321036263
3 0.587859155509931
};
\addlegendentry{FSD50K ($|\mathcal{E}^c|$=50)}
\addplot [semithick, green, opacity=0.7, dash pattern=on 5.55pt off 2.4pt, mark=x, mark size=3, mark options={solid}]
table {%
0 0.918
1 0.95
2 0.964
3 0.968
4 0.970625
5 0.97125
};
\addlegendentry{ESC-50 ($|\mathcal{E}^c|$=20)}
\addplot [semithick, blue, opacity=0.7, forget plot]
table {%
-0.3 0.583
6.3 0.583
};
\addplot [semithick, green, opacity=0.7, forget plot]
table {%
-0.3 0.948
6.3 0.948
};
\addplot [semithick, red, opacity=0.7, forget plot]
table {%
-0.3 0.485
6.3 0.485
};

\nextgroupplot[
scaled y ticks=manual:{}{\pgfmathparse{#1}},
tick align=outside,
tick pos=left,
title={$\text{FS}^{|\mathcal{E}^c|}_{\text{MI+LDA}}$},
x grid style={darkgray176},
xlabel={K},
xmajorgrids,
xmin=0, xmax=6.25,
xtick style={color=black},
xtick={1,2,3,4,5,6},
xticklabels={1,2,4,8,16,32},
y grid style={darkgray176},
ymajorgrids,
ymin=0.35, ymax=1,
ytick style={color=black},
yticklabels={}
]
\addplot [semithick, blue, opacity=0.7, dash pattern=on 5.55pt off 2.4pt, mark=*, mark size=3, mark options={solid}]
table {%
1 0.526
2 0.5625
3 0.5625
4 0.394444444444444
5 0.534166666666666
6 0.554166666666666
};
\addplot [semithick, blue, opacity=0.7, dash pattern=on 5.55pt off 2.4pt, mark=x, mark size=3, mark options={solid}]
table {%
1 0.578869047619047
2 0.611607142857142
3 0.617559523809523
4 0.594642857142857
5 0.454464285714285
6 0.558928571428571
};
\addplot [semithick, blue, opacity=0.7, dash pattern=on 5.55pt off 2.4pt, mark=diamond*, mark size=3, mark options={solid}]
table {%
1 0.59280303030303
2 0.640151515151515
3 0.65
4 0.646590909090909
5 0.634848484848484
6 0.576515151515151
};
\addplot [semithick, green, opacity=0.7, dash pattern=on 5.55pt off 2.4pt, mark=x, mark size=3, mark options={solid}]
table {%
1 0.96625
2 0.969
3 0.9625
4 0.941875
5 0.96625
};
\addplot [semithick, red, opacity=0.7, dash pattern=on 5.55pt off 2.4pt, mark=*, mark size=3, mark options={solid}]
table {%
1 0.5085
2 0.502561971315468
3 0.451533231122268
};
\addplot [semithick, red, opacity=0.7, dash pattern=on 5.55pt off 2.4pt, mark=x, mark size=3, mark options={solid}]
table {%
1 0.551061271817977
2 0.520618815054515
3 0.516155771818713
};
\addplot [semithick, red, opacity=0.7, dash pattern=on 5.55pt off 2.4pt, mark=diamond*, mark size=3, mark options={solid}]
table {%
1 0.558874222458797
2 0.5506071628354
3 0.533389326582581
};
\addplot [semithick, blue, opacity=0.7]
table {%
0 0.583
6.25 0.583
};
\addplot [semithick, green, opacity=0.7]
table {%
0 0.948
6.25 0.948
};
\addplot [semithick, red, opacity=0.7]
table {%
0 0.485
6.25 0.485
};
\end{groupplot}

\end{tikzpicture}
    \caption{Accuracy as we increase $K$. Solid horizontal lines indicate the baseline, CLAP23. The left plot shows continual improvements as we use \acrshort{mi} as a feature selector before averaging, while we see a more complex relationship when using \acrshort{mi} as an intermediary step before \acrshort{lda}.}
    \label{fig:MI_Combined}
\end{figure}

\section{Performance Evaluation}
\label{sec:performance}

We compared different zero-shot and \acrshort{fs} methods to each other using their classification accuracy (BBL and ESC-50) or the \acrshort{map} (FSD50K). A summary of the results is given in Table~\ref{tab:results}. The \acrshort{fs} methods used \acrshort{cos} as the distance metric and non-normalized embeddings. Using the \acrshort{mse} instead of \acrshort{cos} as the distance metric strongly deteriorated the results. Normalizing the embeddings showed limited influence on the results. 

\subsection{Zero-Shot vs. Few-Shot Methods}
A comparison of zero-shot and \acrshort{fs} methods is given in Table~\ref{tab:results}. For the zero-shot methods, CLAP~23 performs best for all three datasets. For the \acrshort{fs} methods, we see increased performance for an increasing $|\mathcal{E}^c|$. Comparing $\text{FS}^{10}_{\text{AVG}}$  to CLAP~23 the accuracy in BBL increases from $0.623$ to $0.678$. For $\text{FS}^{50}_{\text{AVG}}$, the increase is even higher to $0.716$. Consequently, a small number of reference samples per class can be used to improve the performance compared to zero-shot baselines. We use zero-shot baselines as a reference point to assess how our system improves when transitioning from a zero-shot to a few-shot system. For $\text{FS}^{|\mathcal{E}^c|}_{\text{AVG}}$, the performance is always better than the zero-shot baselines. Interestingly, for $\text{FS}^{|\mathcal{E}^c|}_{\text{LDA}}$ the performance only improves for higher $|\mathcal{E}^c|$. For $|\mathcal{E}^c|=10$, the performance is even reduced compared to the zero-shot baselines. 

To investigate the influence of $|\mathcal{E}^c|$ further, we plot the mean and standard deviation of the accuracy for multiple runs using ESC-50 in Figure~\ref{fig:ESC50_var}. For small $|\mathcal{E}^c|$, the standard deviation of the accuracy over multiple runs is higher than for a large $|\mathcal{E}^c|$. This is expected as more audio embeddings reduce the influence of individual files, leading to a more reliable $e_c$.  Increasing $|\mathcal{E}^c|$ leads to an increase in accuracy for all FS methods. The mean-performance of $\text{FS}^{|\mathcal{E}^c|}_{\text{LDA}}$ is always lower than of $\text{FS}^{|\mathcal{E}^c|}_{\text{AVG}}$. We assume that AVG is preferable over LDA here, as the audio embeddings are high-dimensional, and the curse of dimensionality reduces the performance of LDA. Compared to CLAP~23, less than five embeddings per class are required to improve the results. Compared to a CLAP model that is fine-tuned on ESC-50, the FS methods perform slightly worse. Note that fine-tuning comes with higher computational and temporal costs compared to the investigated FS methods and might deteriorate performance on unseen classes.

To assess the effect of zero-shot versus FS methods on a class-by-class comparison, we visualize the confusion matrix for BBL using CLAP~23\footnote{
Some label adjustments were needed to align with CLAP’s mappings, e.g., `low rev engines' $\rightarrow$ `idling tractor' and `baby/kids crying' $\rightarrow$ `baby crying', highlighting the limitations of text embeddings.}   and $\text{FS}^{|\mathcal{E}^c|}_{\text{AVG}}$  in Figure~\ref{fig:Confusion_Matrices}. Overall, accuracy improves, with $\text{FS}^{|\mathcal{E}^c|}_{\text{AVG}}$ outperforming CLAP~23 by approximately ~$10$ percentage points. Notable gains occur in classes that CLAP~23 misclassifies (e.g., `car', `fan/AC') or over-predicts (e.g., `eating'), though some, like `baby/kids crying' or `eating', see slight accuracy drops. Audio-based embeddings help distinguish classes that textual descriptions confuse, such as `car' vs. `car cabin', improving `car' accuracy from nearly 0\% to 45\%. However, class overlap remains challenging, particularly for paired categories like `airplane' and `airplane cabin'.

\subsection{Few-Shot: Dimensionality Reduction using Mutual Information}

We introduce \acrshort{mi} as a preprocessing step of the embeddings to retain only the most relevant features, aiming to improve classification performance while mitigating the curse of dimensionality, particularly for \acrshort{lda}. A comparison of  $\text{FS}^{|\mathcal{E}^c|}_{\text{MI+AVG}}$ and $\text{FS}^{|\mathcal{E}^c|}_{\text{AVG}}$ in Table~\ref{tab:results} shows minor positive effect of MI. For $\text{FS}^{|\mathcal{E}^c|}_{\text{MI+LDA}}$, a positive effect of MI can be seen for $|\mathcal{E}^c| =50$. 

To investigate the number of required features, we plot the accuracy/mAP over $K$ in Figure~\ref{fig:MI_Combined}. For MI+AVG, increasing $K$ improves the performance till it slightly surpasses using all 1024 features. For the largest $K$, around 700 to 800 features are selected. This suggests that most features are relevant to distinguish between audio classes. For MI+LDA, increasing $K$ does not automatically lead to increased performance. LDA has difficulties handling high-dimensional embeddings (curse of dimensionality), such that the performance decreases. These observations are in line with the overall reduced performance of LDA compared to a simple averaging approach. As most features are important, and LDA struggles with high-dimensional data, AVG performs better and more stable. 

\section{Conclusion}
\label{sec:conclusion}
In this study, we improved audio classification by transitioning from zero-shot to few-shot methods, addressing the limitations of noisy text embeddings. Our few-shot approach, utilizing a small number of audio samples, consistently outperformed zero-shot classifiers on different datasets by 2 to 10~\% points in accuracy. This success is due to the more reliable class representations formed by direct audio embeddings, enhancing the system's robustness. 
Future work could explore extending this approach to more complex tasks, such as hierarchical classification or scenarios involving overlapping and concurrent sound events, further advancing the capabilities of few-shot audio classification.
\newpage

\bibliographystyle{IEEEtran}
\bibliography{mybib}

\end{document}